\documentclass{PoS}

\usepackage{amssymb,amsmath,amsbsy}
\usepackage{cite}
\usepackage[utf8x]{inputenc}

\usepackage{float}

\usepackage{graphicx}
\usepackage{psfrag,fancyhdr,epsfig}

\newcommand{\newc}{\newcommand*}

\long\def\begincomment#1\endcomment{%
        \begingroup\sf\baselineskip12pt#1\endgroup}

\newc{\etal}{\textrm{et al.}} 
\newc{\eg}{\textrm{e.g.}} 
\newc{\ie}{\textrm{i.e.}}
\newc{\etc}{\textrm{etc}}
\newc\vs{\textrm{vs.}}
\newc{\cl}{\rm {C.L.}}

\newc{\eV}{\ensuremath{\,\mathrm{eV}}}
\newc{\KeV}{\ensuremath{\,\mathrm{keV}}}
\newc{\MeV}{\ensuremath{\,\mathrm{MeV}}}
\newc{\GeV}{\ensuremath{\,\mathrm{GeV}}}
\newc{\TeV}{\ensuremath{\,\mathrm{TeV}}}

\newc{\invpb}{\ensuremath{/\text{pb}}}
\newc{\invfb}{\ensuremath{\,\text{fb}^{-1}}}
\newc\nb{\ensuremath{\,\mathrm{nb}}} \newc\pb{\ensuremath{\,\mathrm{pb}}} \newc\fb{\ensuremath{\,\mathrm{fb}}}
\newc\pc{\ensuremath{\,\mathrm{pc}}}
\newc\kpc{\ensuremath{\,\mathrm{kpc}}}
\newc\mpc{\ensuremath{\,\mathrm{Mpc}}}
\newc\ps{\ensuremath{\,\mathrm{ps}}} 
\newc\cmeter{\ensuremath{\,\mathrm{cm}}} 
\newc\meter{\ensuremath{\,\mathrm{m}}} 
\newc\kmeter{\ensuremath{\,\mathrm{km}}}
\newc\second{\ensuremath{\,\mathrm{s}}}
\newc\msecond{\ensuremath{\,\mathrm{ms}}}
\newc\nsecond{\ensuremath{\,\mathrm{ns}}}
\newc\psecond{\ensuremath{\,\mathrm{ps}}}

\newc{\D}{\partial}
\newc{\shh}{\lambda_{h}}
\newc{\shf}{\lambda_{h\phi}}
\newc{\sff}{\lambda_{\phi}}
\newc{\sss}{\lambda_{s}}
\newc{\ssisi}{\lambda_{\sigma}}
\newc{\sssi}{\lambda_{s \sigma}}
\newc{\shs}{\lambda_{h s}}
\newc{\shsi}{\lambda_{h \sigma}}
\newc{\sfs}{\lambda_{\phi s}}
\newc{\sfsi}{\lambda_{\phi \sigma}}

\newcommand{\dd}{\mathrm{d}}

\let\alp=\alpha

\newc{\chisqmin}{\ensuremath{\chi^2_{\mathrm{min}}}}
\newc{\Delchisq}{\ensuremath{\Delta\chi^2}}
\newc{\chisq}{\ensuremath{\chi^2}}
\newc{\like}{\ensuremath{\mathcal{L}}}

\newc\lsim{\ensuremath{\mathrel{\rlap{\lower4pt\hbox{\hskip1pt$\sim$}}\raise1pt\hbox{$<$}}}}
\newc\gsim{\ensuremath{\mathrel{\rlap{\lower4pt\hbox{\hskip1pt$\sim$}}\raise1pt\hbox{$>$}}}}
\newc{\VEV}[1]{\ensuremath{\langle #1 \rangle}}
\newc{\dl}{\ensuremath{\stackrel{\leftarrow}{D}}}
\newc{\dr}{\ensuremath{\stackrel{\rightarrow}{D}}}

\newc{\bcenter}{\begin{center}}    \newc{\ecenter}{\end{center}}
\newc{\bfl}{\begin{flushleft}}    \newc{\efl}{\end{flushleft}}
\newc{\bfr}{\begin{flushright}}    \newc{\efr}{\end{flushright}}

\newc{\bi}{\begin{itemize}}
\newc{\ei}{\end{itemize}}
\newc{\bed}{\begin{description}}
\newc{\eed}{\end{description}}
\newc{\ben}{\begin{enumerate}}
\newc{\een}{\end{enumerate}}

\newc{\be}{\begin{equation}}
\newc{\ee}{\end{equation}}
\newc{\ba}{\begin{eqnarray}}
\newc{\ea}{\end{eqnarray}}
\newc{\ra}{\rightarrow}
\def\beq{\begin{equation}}
\def\eeq{\end{equation}}
\def\bea{\begin{eqnarray}}
\def\eea{\end{eqnarray}}
\def\bs#1\es{\begin{split}#1\end{split}}
\def\bsub#1\esub{\begin{subequations}#1\end{subequations}}

\newc{\alphas}{\ensuremath{\alpha_s}}
\newc{\alphatwo}{\ensuremath{\alpha_2}}
\newc{\alphaone}{\ensuremath{\alpha_1}}
\newc{\alphai}[1]{\ensuremath{\alpha_{#1}}}
\newc{\alphaem}{\ensuremath{\alpha_{\mathrm{em}}}}
\newc{\alphaeff}{\ensuremath{\alpha_{\mathrm{eff}}}}
\newc{\sineff}{\ensuremath{\sin \theta_{\mathrm{eff}}}}
\newc{\sinsqeff}{\ensuremath{\sin^2 \theta_{\mathrm{eff}}}}
\newc{\dalphahad}{\ensuremath{\Delta \alpha_{\mathrm{had}}}}
\newc{\yt}{\ensuremath{h_t}} \newc{\yb}{\ensuremath{h_b}} \newc{\ytau}{\ensuremath{h_{\tau}}}
\newc\mz{\ensuremath{M_Z}} 
\newc\mw{\ensuremath{m_W}}
\newc\mZ{\mz}        \newc\mW{\mw}
\newc\mhsm{\ensuremath{ m_{H_{\mathrm{SM}}}}}
\newc{\mtop}{\ensuremath{ m_t}}               \newc{\mtpole}{\ensuremath{ M_t}}
\newc{\mbottom}{\ensuremath{ m_b}} 
\newc{\mtau}{\ensuremath{ m_{\tau}}}
\newc{\mt}{\mtpole}
\newc{\mb}{\mbottom} 
\newc{\rgg}{\ensuremath{R_{h}(\gamma\gamma)}}
\newc{\rzz}{\ensuremath{R_{h}(ZZ)}}
\newc{\rtwogg}{\ensuremath{R_{h_2}(\gamma\gamma)}}
\newc{\rtwozz}{\ensuremath{R_{h_2}(ZZ)}}
\newc{\ronegg}{\ensuremath{R_{h_1}(\gamma\gamma)}}
\newc{\ronezz}{\ensuremath{R_{h_1}(ZZ)}}
\newc{\rsiggg}{\ensuremath{R_{h_\textrm{sig}}(\gamma\gamma)}}
\newc{\rsigzz}{\ensuremath{R_{h_\textrm{sig}}(ZZ)}}
\newc{\llbar}{\ensuremath{\ell\bar{\ell}}}
\newc{\tauptaum}{\ensuremath{ \tau^+\tau^-}}
\newc{\qqbar}{\ensuremath{ q\bar{q}}} \newc{\ppbar}{\ensuremath{ p\bar{p}}}
\newc{\bbbar}{\ensuremath{ b\bar{b}}} \newc{\ttbar}{\ensuremath{ t\bar{t}}}
\newc{\ffbar}{\ensuremath{ f\bar{f}}} \newc{\tautaubar}{\ensuremath{ \tau\bar{\tau}}}
\newc{\mchi}{\ensuremath{m_{\chi}}}
\newc{\squark}{\ensuremath{\tilde{q}}}
\newc{\slepton}{\ensuremath{\tilde{l}}}
\newc{\gluino}{\ensuremath{\tilde{g}}} 
\newc{\wino}{\ensuremath{\tilde{W}}}
\newc{\bino}{\ensuremath{\tilde{B}}}
\newc{\mgluino}{\ensuremath{{m_{\gluino}}}}
\newc{\tone}{\ensuremath{{\tilde{t}_1}}}

\title{Single-field inflation in models with an $R^2$ term}

\ShortTitle{Single-field inflation in models with an $R^2$ term}

\author{Ignatios Antoniadis\\
        LPTHE, Sorbonne Universite, CNRS, 4 Place Jussieu, 75005 Paris, France;\\
		Albert Einstein Center, Institute of Theoretical Physics, University of Bern, Sidlerstrasse 5, CH-3012, Bern, Switzerland\\        
        E-mail: \email{antoniad@lpthe.jussieu.fr}}

\author{\speaker{Alexandros Karam}\\
        National Institute of Chemical Physics and Biophysics, R\"avala 10, 10143 Tallinn, Estonia;\\
        Physics Department, University of Ioannina, GR--45110 Ioannina, Greece\\
        E-mail: \email{alexandros.karam@kbfi.ee}}

\author{Angelos Lykkas\\
        Physics Department, University of Ioannina, GR--45110 Ioannina, Greece\\
        E-mail: \email{alykkas@cc.uoi.gr}}

\author{Thomas Pappas\\
        Physics Department, University of Ioannina, GR--45110 Ioannina, Greece\\
        E-mail: \email{thomasdpappas@gmail.com}}

\author{Kyriakos Tamvakis\\
        Physics Department, University of Ioannina, GR--45110 Ioannina, Greece\\
        E-mail: \email{tamvakis@uoi.gr}}

\abstract{We present two cases where the addition of the $R^2$ term to an inflationary model leads to single-field inflation instead of two-field inflation as is usually the case. In both cases we find that the effect of the $R^2$ term is to reduce the value of the tensor-to-scalar ratio $r$.}

\FullConference{Corfu Summer Institute 2019 "School and Workshops on Elementary Particle Physics and Gravity" (CORFU2019)\\
		31 August - 25 September 2019\\
		Corfù, Greece}

\begin{document}

\section{Introduction}

The theory of the exponential expansion of the Universe, better known as cosmic inflation, was originally proposed as a solution to the horizon and flatness problems~\cite{Guth1981, Linde1982}. It was soon after realized that inflation can also provide a mechanism that explains how primordial inhomogeneities were magnified to cosmic size and became the seeds for the growth of structure in the Universe~\cite{Hawking1982, Starobinsky1982, Guth1982, Linde1983a}. 

The Starobinsky model~\cite{Starobinsky1980} is among the simplest and most successful inflationary models, where an $R^2$ term is added to the Einstein-Hilbert action
\begin{equation}
S_{\rm Star.} = \frac{M^2_{\rm Pl}}{2}\int \mathrm{d}^4x\sqrt{-g} \left( R +\frac{1}{6m^2}R^2\right)\ .
\end{equation}
This model belongs to the general class of $F(R)$ theories~\cite{Capozziello2006, Briscese2007, Nojiri2008, Nojiri2008a, DeFelice2010, Capozziello2011, Clifton2012, Rinaldi2014, Bamba2014, Broy2015a, Odintsov2016, Guzzetti2016, Vernov2019}\footnote{Which are equivalent to the scalar-tensor theories of gravity~\cite{Wagoner1970, Damour1992, Damour1993, Barrow1995, Garcia-Bellido1995a, Boutaleb-Joutei1997, Boisseau2000, Morris2001, Esposito-Farese2001, Chiba2003, Clifton2012, Stabile2013, Chiba2013a, Obukhov2014, Jaerv2015a, Kuusk2016a, Vilson2017, Tambalo2017, Artymowski2017, Bhattacharya2017, Bhattacharya2017a, Burns2016, Karam2017, Karamitsos2017, Karamitsos2018, Karam2018, Karam2019a, Steinwachs2019}.}.
\begin{equation}
S_F = \frac{M^2_{\rm Pl}}{2}\int \mathrm{d}^4x\sqrt{-g} \, F(R)
\end{equation}
which are equivalently described by an action of this form
\begin{equation}
S[g_{\mu\nu},\chi] = \frac{M^2_{\rm Pl}}{2} \int d^{4}x \sqrt{- g}~ \left[ F'(\chi) (R - \chi) + F(\chi) \right]\ .
\end{equation}
after we introduce an auxiliary real scalar field $\chi$. By performing a Weyl rescaling of the metric and a field redifinition we obtain the following action:
\begin{equation}
S[g_{\mu\nu},\varphi]  = \frac{M^2_{\rm Pl}}{2}\int \mathrm{d}^4x\sqrt{-g} R 
 - \int \mathrm{d}^4x \sqrt{-g} \left[ \frac{1}{2}g^{\mu\nu}\partial_{\mu}\varphi\partial_{\nu}\varphi
 + V(\varphi)\right]~,
\end{equation}
where $\chi$ is minimally coupled to gravity, has a canonical kinetic term and a potential given by the general formula
\begin{equation}
V =   \left(\frac{M^2_{\rm Pl}}{2}\right)       \frac{\chi F'(\chi) - F(\chi)}{F'(\chi)^{2}}~, \quad F'(\chi) =  \exp \left(  \sqrt{\frac{2}{3}} \varphi/M_{\rm Pl} \right)~,\quad
  \varphi =  \frac{\sqrt{3}M_{\rm Pl}}{\sqrt{2}}\ln F'(\chi)\ .
\end{equation}
For the $\left( R + R^2 \right)$ model we have
\begin{equation} 
V(\varphi) = \frac{3}{4} M^2_{\rm Pl}m^2\left[ 1- \exp\left(-\sqrt{\frac{2}{3}}\varphi/M_{\rm Pl}\right)\right]^2~.
\end{equation}

Recently, the Starobinksy model has been combined with other popular models such as Higgs inflation~\cite{Artymowski2015, Bruck2015, Asaka2016, Artymowski2016, Kaneda2016, Calmet2016, Bruck2016a, Wang2017, Ema2017, Mori2017, Pi2018, He2018, Gorbunov2018, Ghilencea2018, Wang2018a, Antoniadis2018, Gundhi2018, Enckell2018a, He2019, Bezrukov2019, Canko2019, Samart2019, Ema2019}. As a result, there are two fields that can play the role of the inflaton and the analysis becomes more complicated.

\begin{figure}[H]
\centering
\includegraphics[width=\textwidth]{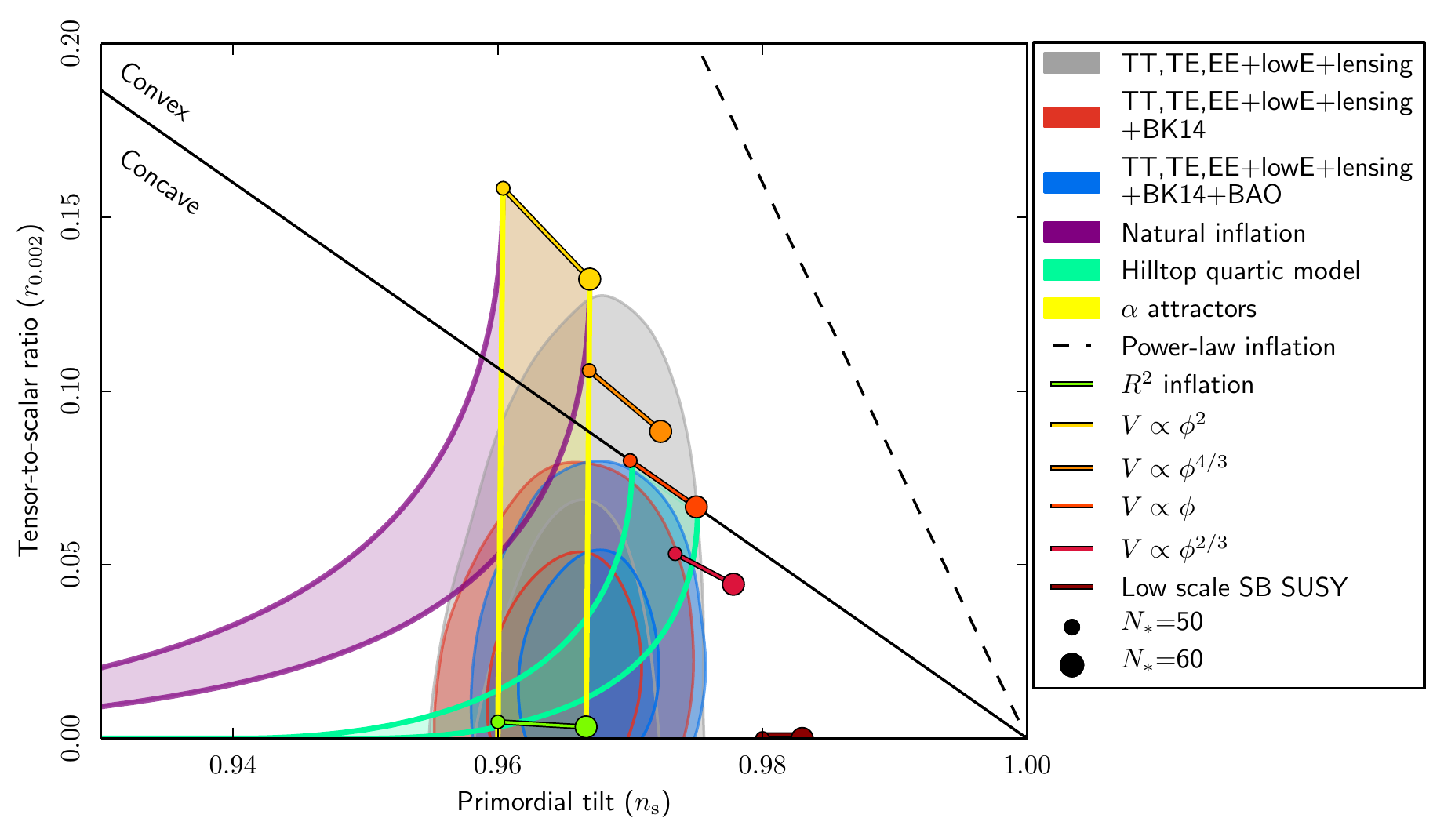}
\caption{\sf Marginalized joint 68\% and 95\% CL regions for $n_S$ and $r$ at $k = 0.002 \mathrm{Mpc}^{−1}$ from Planck 2018~\cite{Akrami2018} compared to the theoretical predictions of selected inflationary models.}
\label{fig:Planck2018}
\end{figure}

The 2018 results from the Planck satellite~\cite{Akrami2018} in the $n_S$ versus $r$ plane are shown in Fig.~\ref{fig:Planck2018}

The simplest models like $\phi^4$, $\phi^3$ and $\phi^2$ have already been ruled out by the Planck $2015$ data~\cite{Ade2015}, while a little more convoluted models, such as the $\alpha$-attractors~\cite{Kallosh2013b}, the Starobinsky~\cite{Starobinsky1980} and some non-minimally coupled models such as Higgs inflation~\cite{Bezrukov2008, Bezrukov2009} yield predictions that still comply with the observations.

\section{Non-minimal Coleman-Weinberg inflation with an $R^2$ term}
\label{sec:CWSI}

In~\cite{Karam2019}, we considered the non-minimal Coleman-Weinberg model where the Planck mass is dynamically generated through the vacuum expectation value (VEV) of a real scalar field $\phi$, plus an $R^2$ term\footnote{See also~\cite{Cooper1981, Shaposhnikov2009a, Garcia-Bellido2011, Khoze2013, Bezrukov2013b, Gabrielli2014, Salvio2014, Csaki2014, Kannike2015, Barvinsky2015, Marzola2016, Barrie2016, Marzola2016b, Rinaldi2016a, Farzinnia2016, Kannike2016, Karananas2016, Ferreira2016, Tambalo2017, Kannike2017, Artymowski2017, Racioppi2017, Ferreira2017, Salvio2017, Karam2017, Kannike2017a, Barnaveli2018, Ferreira2018, Racioppi2018, Karam2018a, Casas2019, Wetterich2019, Oda2019, Tang2019, Kubo2019, Ghilencea2019, Salvio2019, Vicentini2019, Ferreira2019a, Ferreira2019} for other models based on scale invariance.}. The action in the Jordan frame has the form
\begin{equation}
S^{J}=\int \mathrm{d}^4x \sqrt{-\bar{g}} \left[\frac{\xi \phi^2 }{2}\bar{R}+\frac{\alp}{2} \bar{R}^2-\frac{1}{2} \bar{\nabla}^{\mu} \phi \bar{\nabla}_{\mu} \phi-\frac{\lambda_{\phi}}{4}\phi^4 \right]\ .
\end{equation}
We can eliminate the $R^2$ term by introducing an auxiliary real scalar field $\chi$
\begin{equation}
S^{J}=\int \mathrm{d}^4x \sqrt{-\bar{g}} \left[\frac{1}{2}\left(\xi \phi^2+\alpha\chi^2\right)\bar{R}-\frac{\alpha}{8}\chi^4-\frac{1}{2} \bar{\nabla}^{\mu} \phi \bar{\nabla}_{\mu} \phi-\frac{\lambda_{\phi}}{4}\phi^4 \right]\,.
\end{equation}
Then, by performing a Weyl rescaling of the metric (where we introduced a new field $\zeta$
\begin{equation}
g_{\mu \nu}= \Omega^2 \bar{g}_{\mu \nu}\,, \qquad \Omega^2=\left(\alpha\chi^2+\xi\phi^2\right)/M_{\rm Pl}^2 \equiv \frac{\zeta^2}{6M_{\rm Pl}^2} \,,
\end{equation}
the Einstein frame action takes the form
\begin{equation}
S^{E}=\int \mathrm{d}^4x \sqrt{-g}\left[ \frac{M_{\rm Pl}^2}{2} R-\frac{6 M_{\rm Pl}^2}{\zeta^2} \left(\frac{1}{2} \nabla^{\mu} \phi \nabla_{\mu} \phi+\frac{1}{2} \nabla^{\mu} \zeta \nabla_{\mu} \zeta \right)-V^{(0)E}(\phi, \zeta) \right] \,,
\end{equation}
with the tree-level potential given by the formula
\begin{equation}
V^{(0)E}(\phi, \zeta)= \frac{36 M_{\rm Pl}^4}{\zeta^4} \left[ \frac{\lambda_{\phi}}{4} \phi^4 + \frac{1}{8 \alp} \left( \frac{\zeta^2}{6}-\xi \phi^2 \right)^2 \right]\, .
\end{equation}
Note that there are two fields in the potential, $\phi$ and $\zeta$. Usually, scale-invariant beyond the Standard Model theories with multiple scalar fields are studied with the help of the Gildener-Weinberg formalism~\cite{Gildener1976a}.

In this approach, the perturbative minimization happens in two steps. First, the tree-level potential is minimized with respect to its field content. This minimization takes place at a specific energy scale, due to the running of the couplings in the full quantum theory, and defines a flat direction among the scalar fields. Then, one computes the one-loop corrections only along this flat direction, since this is where the corrections play the dominant role, remove the flatness, and determine the physical vacuum.

Now, the tree-level minimization conditions in our theory give this relation:
\begin{equation}
\frac{\dd V^{(0)E}}{\dd \phi} \Big|_{\phi = v_\phi} = \frac{\dd \,V^{(0)E}}{\dd \zeta} \Big|_{\zeta = v_\zeta} = 0 \ , \qquad \Rightarrow \qquad v_{\phi}^2 = \frac{\xi}{6\left( \xi^2 + 2 \alpha \lambda_\phi \right)} v_{\zeta}^2 \ .
\end{equation}
By using an orthogonal rotation of the form
\begin{equation}
\left(\begin{array}{c}
\phi \\
\zeta 
\end{array}
\right)
=  
\left( \begin{array}{cc}
\cos\omega  & -\sin\omega   \\
\sin\omega   & \cos\omega 
\end{array}
\right)
\left(
\begin{array}{c}
s \\
\sigma 
\end{array}
\right)\,, 
\end{equation}
we can diagonalize the mass matrix of the two scalars and we parametrize the mixing angle in this way
\begin{equation}
\tan\omega = \frac{v_\zeta}{v_\phi} \,, 
\qquad  
v_s = \frac{v_\phi}{\cos\omega} = \frac{v_\zeta}{\sin\omega} \ ,
\end{equation}
with respect to the VEVs. The tree-level mass eigenstates are
\begin{equation}
m_{s}^2 = 0 \ , \qquad m_{\sigma}^2=\frac{\xi\left(\xi+12\lambda_{\phi} \alpha+6\xi^2\right)}{6\alpha\left(2\lambda_{\phi}\alpha+\xi^2\right)} M^2_{\rm Pl} \ .
\end{equation}
Notice that the mass eigenvalue of $s$ is exactly zero at tree level since it is the pseudo-Goldstone boson of broken classical scale invariance, while $\sigma$ is the orthogonal eigenstate.

\begin{figure}
\centering
\includegraphics[width=0.65\textwidth]{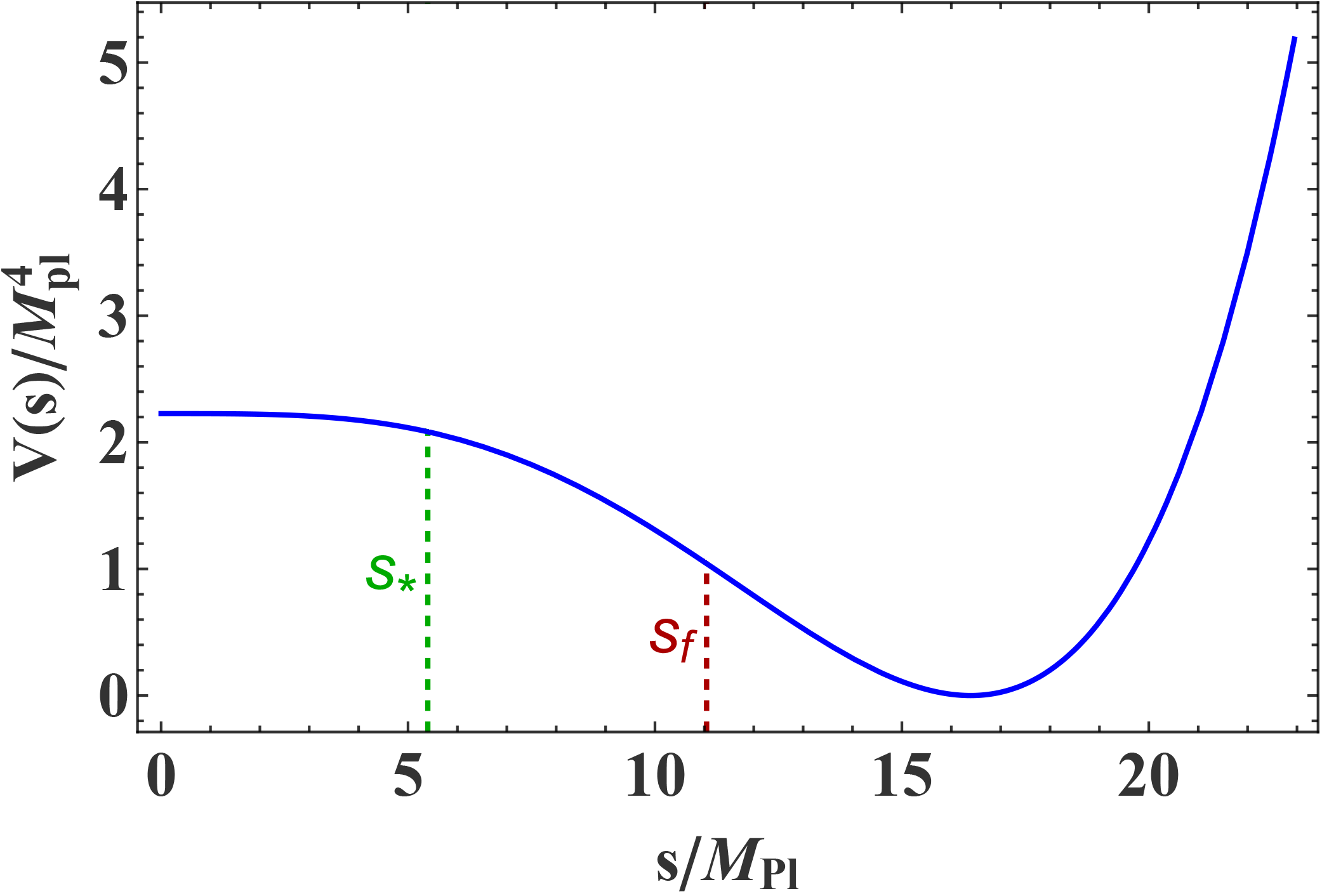}
\caption{\sf An inflaton excursion that yields $N=59.3$ when $\sin\omega= 0.15$. Inflation starts(ends) at the green(red)-dashed line. The values of the observables in this case are $(r,n_s)=(0.014,0.963)$.}
\label{fig:potential}
\end{figure}

The $1$-loop correction along the flat direction takes the form 
\begin{equation}
V^{(1)}=\frac{m^4_{\sigma}}{64 \pi^2 v_s^4} s^4\left[ \log \left(\frac{s^2}{v_s^2}\right)-\frac{1}{2} \right] \,, \qquad v_s^2 = v_{\phi}^2+v_{\zeta}^2 \,, \qquad v^2_\zeta = 6 M^2_{\rm Pl} \ .
\end{equation}
At this point we demand that the full one-loop potential vanishes at the minimum\footnote{This implies that the cosmological constant is zero at the $1$-loop level.}
\begin{equation}
V(v_s) \equiv  V^{(0)E}(v_s) + V^{(1)}(v_s) = 0 \ .
\end{equation}
Finally, we arrive at the following expression for the effective potential along the flat direction
\begin{equation}
V(s)=\frac{m^4_{\sigma}}{128 \pi^2 }\left[ \frac{\sin^2\omega}{36 M_{\rm Pl}^4} s^4 \left(2 \ln \left[ \frac{s^2 \sin^2\omega}{6 M_{\rm Pl}^2} \right]-1 \right)+1  \right] \,, 
\label{eq:1looppot}
\end{equation}
which is shown in Fig.~\ref{fig:potential}.
From \eqref{eq:1looppot} we obtain the radiatively generated mass of the $s$ boson
\begin{equation}
\qquad m_s^2 = \frac{\sin^2\omega}{48 \pi^2}\frac{m^4_\sigma}{M_{\rm Pl}^2} \ .
\end{equation}
Notice that it is loop-suppressed with respect to the mass of $\sigma$. This means that the orthogonal state $\sigma$ effectively decouples and only $s$ plays the role of the inflaton along the flat direction.

\begin{figure}
\centering
\includegraphics[width=0.65\textwidth]{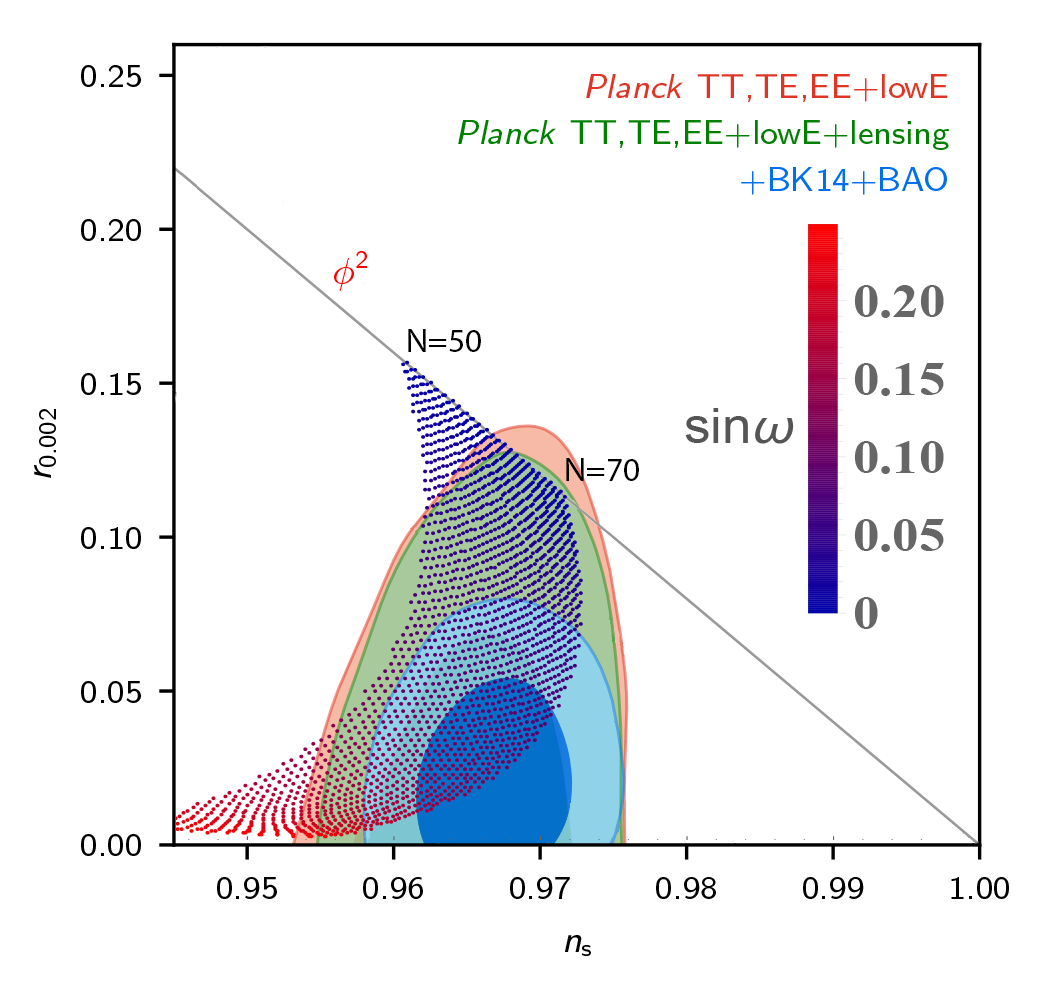}
\caption{\sf The predictions of the model \eqref{eq:1looppot} for a wide range of values for the mixing angle $\omega$ and for field excursions that yield sufficient amount of inflation i.e. 50 to 70 e-folds.}
\label{fig:r-ns_plot}
\end{figure}

The predictions of the model in the $n_s - r$ plane are shown in Fig.~\ref{fig:r-ns_plot}, overlayed with the latest Planck constraints. In the limit of zero mixing angle the potential behaves like the quadratic one, while for larger values we obtain a beautiful banana shape.

\section{Palatini inflation in models with an $R^2$ term}

There are two variational principles that one can apply to the Einstein-Hilbert action in order to derive Einstein's equations: the standard metric variation where the metric is the only dynamical degree of freedom and the connection is the Levi-Civita
\begin{equation}
S = \int \dd^4 x \sqrt{-g} \left( \frac{1 + \xi \phi^2}{2} g^{\mu\nu} R_{\mu\nu} \left( g, \partial g, \partial^2 g \right) - \frac{1}{2} g^{\mu\nu} \nabla_\mu \phi \nabla_\nu \phi - V(\phi) \right) \ ,
\end{equation}
and the Palatini variation\footnote{See~\cite{Sotiriou2010} for a review and~\cite{Bauer2008, Borunda2008, Olmo2011, Bauer2011, Tamanini2011, Enqvist2012, Borowiec2012, Racioppi2017, Rasanen2017, Fu2017, Stachowski2017, Szydlowski2017, Tenkanen2017, Markkanen2018, Carrilho2018, Enckell2018, Enckell2019, Kozak2018, Jaerv2018, Wang2018, Wu2018, Szydlowski2018, Antoniadis2018, Rasanen2018a, Rasanen2018, Almeida2018, Rubio2019, Shimada2019, Takahashi2019, Jinno2019a, Tenkanen2019a, Edery2019, Jinno2019, Aoki2019, Giovannini2019, Bostan2019a, Bostan2019, Tenkanen2019, Gialamas2019, Racioppi2019} for various applications.} where the metric and the connection are assumed to be independent variables and one varies the action with respect to both of them
\begin{equation}
S = \int \dd^4 x \sqrt{-g} \left( \frac{1 + \xi \phi^2}{2} g^{\mu\nu} R_{\mu\nu} \left( \Gamma, \partial \Gamma \right) - \frac{1}{2} g^{\mu\nu} \nabla_\mu \phi \nabla_\nu \phi - V(\phi) \right) \ .
\end{equation}
Note that even though both variational principles lead to the same field equation for an action whose Lagrangian is linear in $R$ and is minimally coupled, this is no longer true for a more general action.

In~\cite{Antoniadis2018, Antoniadis2019}, we considered a real scalar field $\phi$, in general non-minimally coupled to gravity plus an $R^2$ term in the Palatini formalism\footnote{See also~\cite{Enckell2019} where the authors consider a similar setting.}
\begin{equation}
S=\int\dd^4x\sqrt{-g}\left\{\frac{1}{2}\left(M_0^2+\xi\phi^2\right)R+\frac{\alpha}{4}R^2-\frac{1}{2}\left(\nabla\phi\right)^2-V(\phi)\right\}, \quad R = g^{\mu\nu} R^\rho_{\ \mu\rho\nu} \left( \Gamma, \partial \Gamma \right) \ .
\end{equation}
Introducing an auxiliary scalar $\chi$ we eliminate the $R^2$ term 
\begin{equation}
S\,=\,\int\,\dd^4x\,\sqrt{-g}\left\{\,\frac{1}{2}\left(M_0^2+\,\alpha\chi^2+\xi\phi^2\right)R\,-\frac{1}{2}\left(\nabla\phi\right)^2-\frac{\alpha}{4}\chi^4\,-V(\phi)\,\right\} \ ,
\end{equation}
and by performing a Weyl rescaling of the metric 
\begin{equation}
\bar{g}_{\mu\nu}\,=\,\Omega^2(\phi)\,g_{\mu\nu}\qquad{\text{with}}\qquad\Omega^2(\phi)\,=\,\frac{M_0^2+\xi\phi^2+\alpha\chi^2}{M_P^2}\,,
\end{equation}
we bring the action to this simpler form:
\begin{equation}
S\,=\,\int\,\mathrm{d}^4x \sqrt{-\bar{g}}\left\{\,\frac{1}{2}M_P^2\,\bar{R}\,-\frac{1}{2}\frac{\left(\bar{\nabla}\phi\right)^2}{\Omega^2}\,-\bar{V}\right\}\ , \quad \bar{V}(\phi,\chi)\,=\,\frac{1}{\Omega^4}\left(V(\phi)+\frac{\alpha}{4}\chi^4\right)\,.
\end{equation}
Notice that no kinetic term has been generated for $\chi$, which is usually called the scalaron in the metric formalism, therefore its equation of motion reduces to just a constraint. This means that $\phi$ is the only propagating degree of freedom which can play the role of the inflaton, contrary to the metric case where we would have two-field inflation.

Using the constraint equation for $\chi$
\begin{equation}
\delta_\chi S = 0 \rightarrow \chi^2\,=\,\frac{\frac{4V(\phi)}{(M_0^2+\xi\phi^2)}\,+\,\frac{(\nabla\phi)^2}{M_P^2}}{\left[1-\frac{\alpha(\nabla\phi)^2}{M_P^2(M_0^2+\xi\phi^2)}\right]}
\end{equation}
we can eliminate it altogether and arrive at the action
\begin{equation}
S\approx\int\dd^4x\,\sqrt{-g}\left\{\,\frac{M_P^2}{2}R\,-\frac{1}{2}\frac{(\nabla\phi)^2}{\Omega_0^2}\left(\frac{1}{1+\frac{4\tilde{\alpha} V}{\Omega_0^4}}\right)\,-\frac{V}{\Omega_0^4}\left(\frac{1}{1+\frac{4\tilde{\alpha} V}{\Omega_0^4}}\right)\,+\,\mathcal{O}\left((\nabla\phi)^4\right)\right\}\ ,
\end{equation}
where $\tilde{\alpha}=\alpha/M_P^4$ and $\Omega_0^2=\left(\,M_0^2\,+\,\xi\phi^2\,\right)/M^2_P$.

The effect of the $R^2$ term is that it decreases the height of the effective potential, which always has a plateau at high field values. Of course, one should also take into account that the rate of change of the field is modified and should perform a field redefinition in order to bring the kinetic term into its canonical form. 

\begin{figure}
\centering
\includegraphics[width=0.65\textwidth]{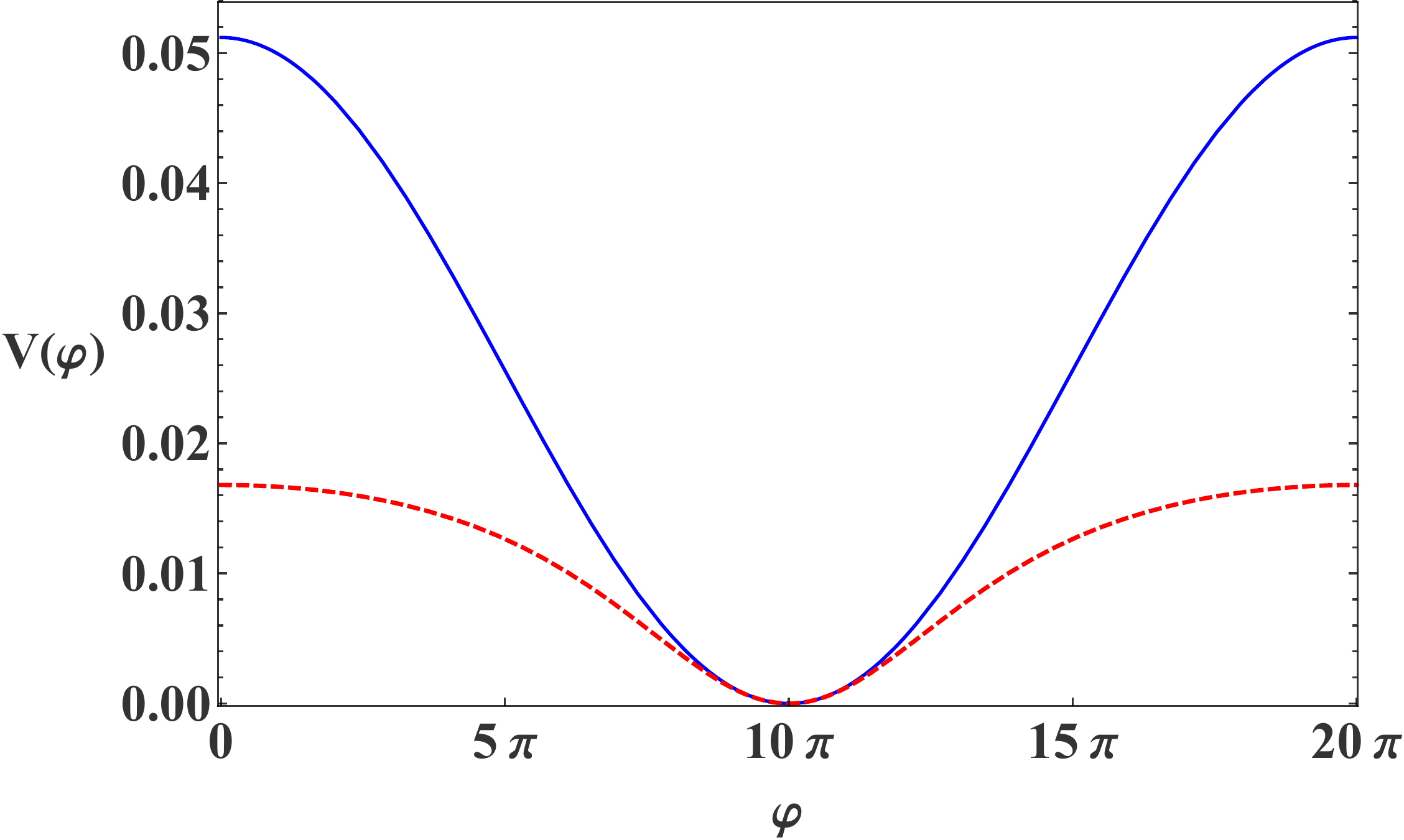}
\caption{\sf Blue curve: The original natural inflation potential $V(\phi)$. Red dashed curve: The potential $\bar{V}(\zeta(\phi))$ in the Palatini formalism. We have chosen $M = 0.4$, $\alpha = 10$, and $f = 10$ (in natural units).}
\label{fig:Natural_flattening}
\end{figure}

\begin{figure}
\centering
\includegraphics[width=0.65\textwidth]{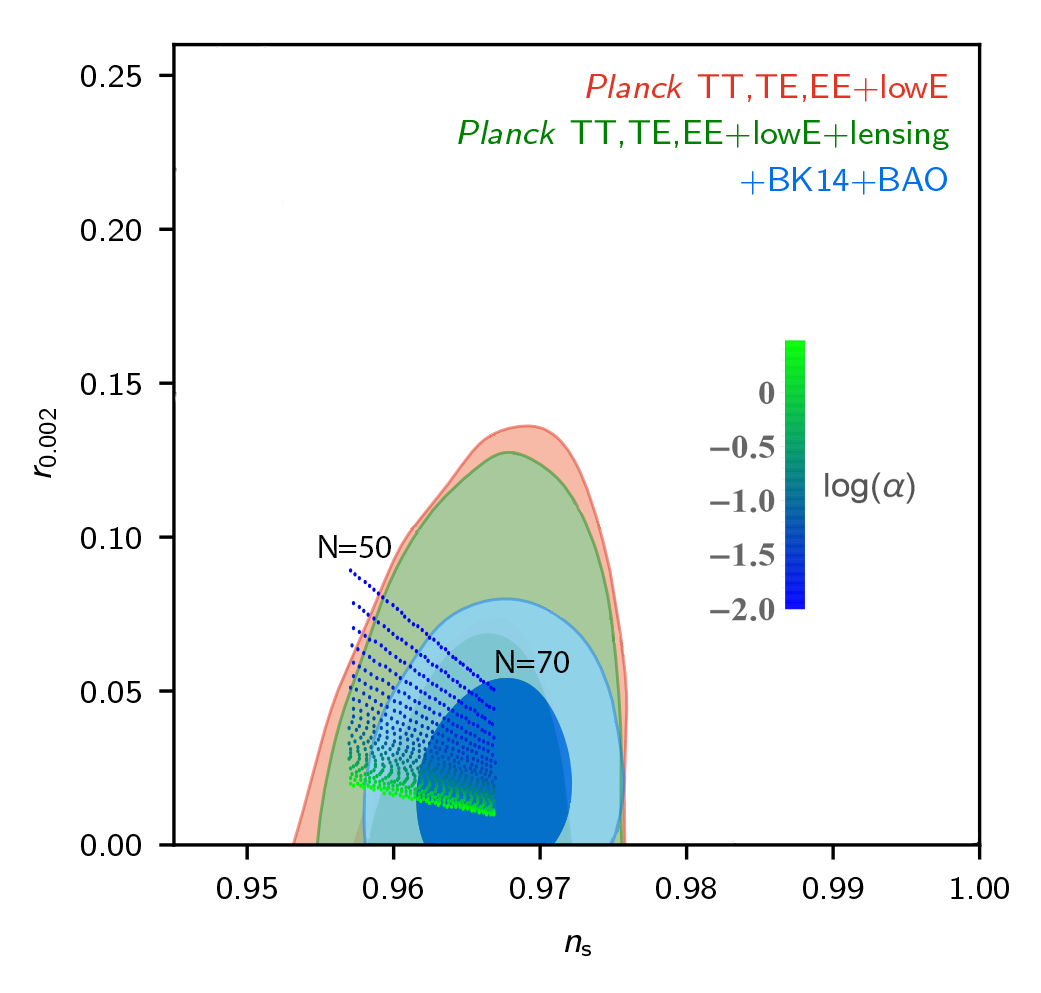}
\caption{\sf $r-n_s$ plot for $M=0.7, f=7$ and $\alpha \in [0.01,3]$ in natural units.}
\label{fig:natural_rns}
\end{figure}

\begin{figure}
\centering
\includegraphics[width=0.65\textwidth]{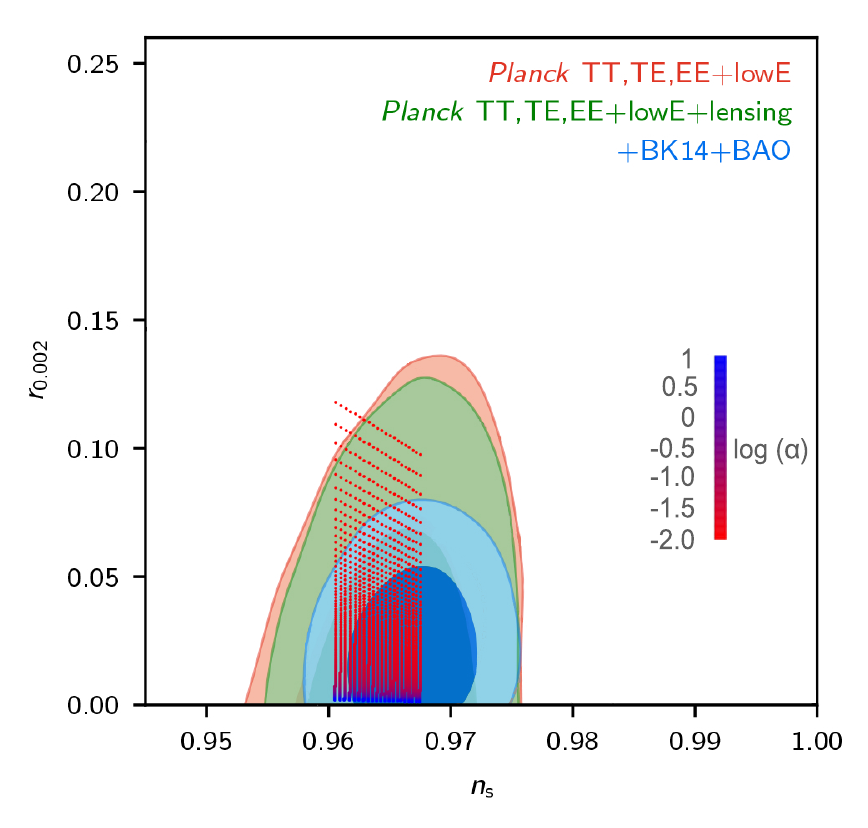}
\caption{\sf The predictions for the inflationary observables in the $n_S-r$ plane for the minimally-coupled quadratic model for $N=50-60$ e-folds. We have set $m=0.1$ and have varied $\alpha$ between $0.01$ and $10$.}
\label{fig:r_ns-squared-m-01}
\end{figure}

\begin{figure}
\centering
\includegraphics[width=.65\textwidth]{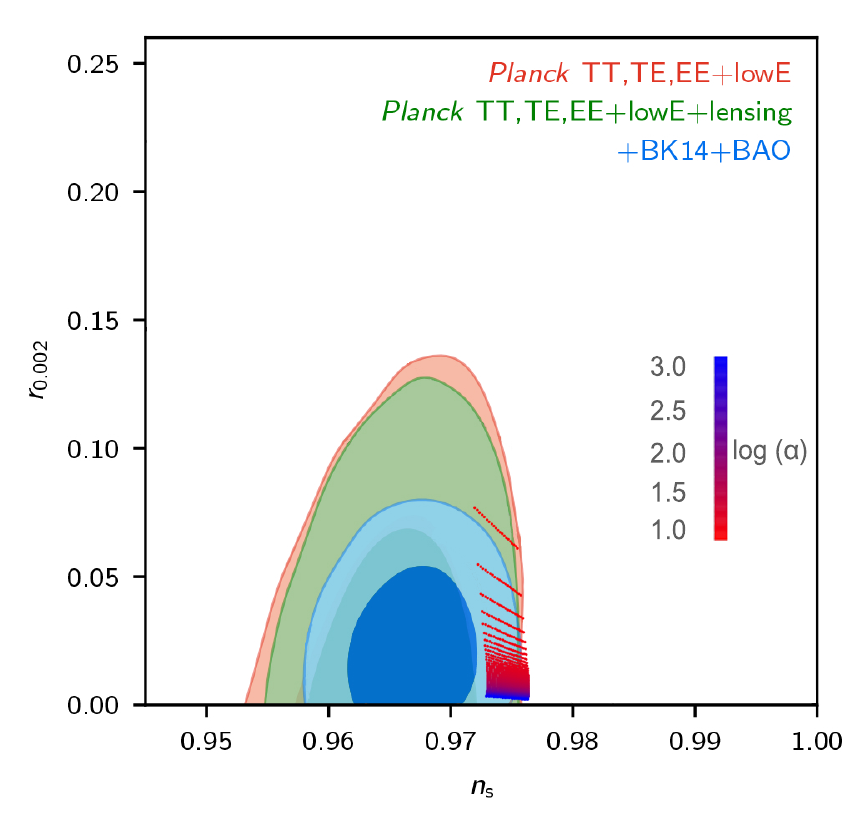}[H]
\caption{\sf The predictions for the inflationary observables in the $n_S-r$ plane for the nonminimal Coleman-Weinberg model and for $N=50-60$ e-folds. We have set $\Lambda = 0.1$ and $\xi = 0.1$ (in Planck units).}
\label{fig:r_ns-cw}
\end{figure}

As a first example, let us consider natural inflation, given by the potential~~\cite{Freese1990, Adams1993}
\begin{equation}
V(\phi) = M^4 \left( 1 + \cos\left( \frac{\phi}{f} \right) \right) \ .
\end{equation}
In Fig.~\ref{fig:Natural_flattening}, the blue curve is the potential in the metric formalism. The red dashed curve is the effective potential with the $R^2$ in the Palatini formalism. As can be seen, the potential has a lower height and has been flattened.
In practice, this means that we can obtain lower values for the tensor-to-scalar ratio $r$ (see Fig.~\ref{fig:natural_rns}) and essentially save some models that were previously excluded by the Planck constraints.
Similarly, for minimal quadratic inflation,
\begin{equation}
V(\phi) = \frac{1}{2} m^2 \phi^2
\end{equation}
which was excluded, we find that relatively small values of $\alpha$ can significantly lower the tensor-to-scalar ratio $r$ (see Fig.~\ref{fig:r_ns-squared-m-01})\footnote{However, accordance with the measured value of power spectrum requires larger values for $\alpha$~\cite{Tenkanen2019a}.}.
The same holds for non-minimally coupled models. For example, in the model considered in~\cite{Kannike2016b}, 
\begin{equation}
S=\int\dd^4x\sqrt{-g}\left\{\frac{1}{2}\left(\xi\phi^2+\alpha\chi^2\right)R\,-\frac{1}{2}\left(\nabla\phi\right)^2\,-\frac{\lambda(\phi)}{4}\phi^4-\frac{\alpha}{4}\chi^4+\Lambda^4\right\} \ ,
\end{equation}
\begin{equation}
V_1(\phi)=\frac{1}{4}\lambda(\phi)\phi^4 + \Lambda^4=\Lambda^4\left(1+\frac{\xi^2\phi^4}{M_P^4}\left(2\ln(\xi\phi^2/M_P^2)-1\right)\right)\,,\quad \langle\phi\rangle^2 = M^2_{\rm P}/\xi
\end{equation}
where the Planck scale is dynamically generated through the Coleman-Weinberg mechanism, depending on the value of the non-minimal coupling $\xi$, interpolates between the quadratic and linear inflation limits, 
both of which are now excluded by Planck. However, in the Palatini formalism with the $R^2$ term, the model can be easily made viable (see Fig.~\ref{fig:r_ns-cw}).


\section{Conclusions}

In conclusion, while in the metric formalism adding an $R^2$ term to our action results in two-field (an exception being the Coleman-Weinberg model we considered in Sec.~\ref{sec:CWSI} using the Gildener-Weinberg formalism), the same does not hold in the Palatini formalism. We obtain a single-field effective potential which is asymptotically flat and has a lower value, which means that we can have a smaller tensor-to-scalar ratio.

\section*{Acknowledgments}
A.K. and T.P. acknowledge support from the Operational Program ``Human Resources Development, Education and Lifelong Learning" which is co-financed by the European Union (European Social Fund) and Greek national funds. A.K. also acknowledges the support of the Estonian Research Council grant MOBJD381 and the ERDF Centre of Excellence project TK133 and also thanks the COST action for the financial support and the organizers of the “Workshop on the Standard Model and Beyond", Corfu 2019, for the hospitality during his stay, and for giving him the opportunity to present this work.

\bibliography{References}{}
\bibliographystyle{utphys}

%

\end{document}